\documentclass[twocolumn,preprintnumbers,nofootinbib,prd]{revtex4}

\usepackage{graphicx}

\usepackage[hypertex]{hyperref}

\newcommand{\lsim}{\lesssim}
\newcommand{\ord}[1]{\mathcal{O}{(#1)}}

\newcommand{\gsim}{\gtrsim}
\newcommand{\beq}{\begin{equation}}
\newcommand{\eeq}{\end{equation}}

\newcommand{\zeff}{Z_{\rm eff}}
\newcommand{\eps}{\varepsilon}
\newcommand{\aae}{\alpha_{ae}}

\begin{document}

\pagestyle{plain}

\title{Thermal Geo-axions}

\author{Hooman Davoudiasl
\footnote{email: hooman@bnl.gov}
}
\affiliation{Department of Physics, Brookhaven National Laboratory,
Upton, NY 11973, USA}

\author{Patrick Huber
\footnote{email: pahuber@vt.edu}
}

\affiliation{Department of Physics, IPNAS, Virginia Tech, Blacksburg, VA
  24061, USA}


\begin{abstract}

  We estimate the production rate of axion-type particles in the core
  of the Earth, at a temperature $T\approx 5000$~K.  We constrain {\it
  thermal geo-axion} emission by demanding a core-cooling rate less than
  $\ord{100}$~K/Gyr, as suggested by geophysics.
  This yields a ``quasi-vacuum" (unaffected by extreme stellar conditions) bound
  on the axion-electron fine structure constant $\aae^{QV} \lsim
  10^{-18}$, stronger than the existing accelerator (vacuum) bound by 4 orders of
  magnitude.  We consider the prospects for measuring the geo-axion
  flux through conversion into photons in a {\it geoscope}; such
  measurements can further constrain $\aae^{QV}$.
\end{abstract}
\maketitle


A variety of scenarios for physics beyond the Standard Model (SM) give
rise to light pseudo-scalar particles, generically referred to as
axions.  The Peccei-Quinn (PQ) solution to the SM strong CP problem
provided the initial context for axions~\cite{Peccei:1977hh}.
Axion-type particles are ubiquitous in string theory constructs and
have also been considered in cosmological model
building~\cite{Preskill:1982cy}.  There are stringent astrophysical
and cosmological constraints on the couplings of axions, as a result
of which they are largely assumed to be very weakly interacting.  Some
of the strongest bounds on axion-SM couplings come from astrophysics,
where stellar evolution and cooling arguments imply that the axion
(PQ) scale $f \gsim 10^9$~GeV.  Such analyzes are based on the
requirement that new exotic processes should not significantly perturb
a standard picture of the energetics that govern the evolution of
various astrophysical objects.  Since axions (or other light weakly
interacting particles) can directly drain energy out of such objects,
one can obtain bounds on the coupling of axions to matter. For a
concise summary of various astrophysical bounds, see
Ref.~\cite{Amsler:2008zzb}.  More recent astrophysical bounds
on axion-type particles have been presented in
Ref.~\cite{Gondolo:2008dd}.

In this work, we consider the possibility that the hot core of the
Earth can convert some of its thermal energy into a flux of axions
of $\ord{{\rm eV}}$ energy
\cite{Sivaram}\footnote{Non-thermal geo-axions, produced in radioactive
decays within the Earth, have been examined
in Ref.~\cite{Liolios:2007gu}. This work does not find a currently
detectable signal even in the most favorable case considered therein.}.
Then, it would be interesting to find out what
bounds can be obtained from geological considerations and also to
determine the prospects for discovering the geo-axions emanating
from the terrestrial core.  The Earth's core is at a
temperature of around $5000\,\mathrm{K}$ corresponding to $0.4\,\mathrm{eV}$.
Although this is a much lower temperature than those of stellar
interiors, which have temperatures of order $\mathrm{keV}$, there
are a number of considerations that motivate our analysis.

First of all, the core of the Earth is only a short distance away,
compared to any astronomical object.  This greatly enhances the
prospects for measuring a geo-axion flux and can potentially
compensate for the low core temperature. Secondly, the Earth's core is
quite different from other axion emitting environments, being mainly
made up of hot molten or crystallized iron.  Hence, in principle, the
intuition and calculations that apply to stellar plasmas may not be
adequate to estimate geo-axion emission and new effects may need to be
considered.  Finally, the Earth's center is a far less extreme medium
compared to stellar media.  The possibility of the dependence of axion
properties on the environment has been proposed~\cite{medium1} in the
context of reports of large vacuum birefringence by
PVLAS~\cite{Zavattini:2005tm}. This result if it were confirmed would
have implied an axion like particle with an axion-photon coupling in
stark violation of astrophysical bounds. As a consequence, a number of
models were developed to reconcile the laboratory result with the
astrophysical bounds~\cite{medium2}. Although the initial PVLAS
result could not be reproduced~\cite{Cantatore:2008zz}, it highlighted
the necessity for obtaining complementary bounds on axions in a wide
variety of production environments.  If axion couplings are
temperature and/or density dependent, the geo-axion bounds could be
viewed as independent new data on axion physics in ``quasi-vacuum"
conditions. Thus, in this letter we do not attempt to supersede
existing, stringent astrophysical bounds, but to supplement them by
examining axion production in a novel environment.

Motivated by the above discussion, we will next derive an estimate for
the thermal geo-axion flux.  We will use geodynamical considerations to
constrain this flux and hence the axion-electron coupling $\aae$ in
the core.  This bound is not competitive with its astrophysical
counterparts, but, as mentioned before, is derived in a very different
regime.  Note that collider bounds on $\aae$ that are derived in a
similar regime are much weaker than our geo-axion bound.  We will then
consider detection of the geo-axion flux, via magnetic conversion into
photons, using a ``geoscope," in analogy
with the helioscope concept \cite{Sikivie:1983ip,vanBibber:1988ge}.
A discussion and a summary of our
results are presented at the end of this work.

The core of the Earth is mainly made of iron (Fe).  The inner core,
which extends to a radius of $R_{ic} \approx 1200$~km, is thought to
be in solid crystalline form at a temperature $T\sim 6000$~K.  The
outer core, which extends to $R_c \approx 3500$~km, is made up of
molten iron at $T\sim 4000$~K~\cite{core}.  Since Fe is a transition
metal, with the electronic configuration $[{\rm Ar}]\; 3d^6\; 4s^2$,
both $3 d$ and $4 s$ electrons are important in determining its
properties.  However, for a simplified treatment, we only consider the
$4 s$ electrons as nearly free.  The effective nuclear charge seen by
the $4 s$ electrons is $\zeff \simeq 5.4$ \cite{Zeff}.

Given that the solid iron core makes up a negligible mass of the total
core, we will ignore its contributions to our estimate.  This is
partly done to avoid a complicated treatment of the interactions of
electrons and phonons inside a hot crystal, far from the plasma
regime.  However, we note that a more complete analysis should take
these effects into account.  We adopt $T_c \approx 5000$~K$\approx
0.4$~eV~\cite{core} as the mean temperature of the molten iron
core.  We are also ignoring the contribution of other trace elements,
such as nickel, which have more or less the same properties as iron,
for our purposes.  Given the metallic nature of the core, we will
treat it as a plasma composed of a degenerate gas of free electrons,
with a Fermi energy $E_F \approx 10.3$~eV \cite{FerroFe}.
The resulting Fermi
momentum is given by $p_F = \sqrt{2 m_e\,E_F }\approx 3.3$~keV,
where $m_e \simeq 0.5$~MeV is the mass of the electron.
These free electrons move
in the background of Fe ions with effective charge $\zeff \simeq 5.4$.
The free electron density in the core is given by \cite{GR}
\beq
n_e = \frac{p_F^3}{3 \pi^2}
\label{ne}
\eeq and hence we get $n_e \approx 2\times 10^{23}$~cm$^{-3}$.  Let us
define the radius \beq a_e = \left( \frac{3}{4\pi n_e}\right)^{1/3},
\label{ae}
\eeq
for the mobile charged particles in the plasma, which we take to
be electrons here.  The quantity
\beq
\Gamma \equiv \frac{\zeff^2 \alpha}{a_e T_c},
\label{Gamma}
\eeq with $\alpha = 1/137$, is a measure of the relative strength of
Coulomb interactions and the kinetic energy of the electrons.  For the
core parameters, we get $a_e \approx 10^{-8}$~cm and $\Gamma \sim
10^3$.  We take $\Gamma \gg 1$ as
indicative of a strongly coupled plasma~\cite{GR}.
Since the iron core of the Earth is in a molten state and
not yet a crystal, this interpretation is reasonable,
despite the large value of $\Gamma$. The effect of the
geomagnetic field in the core on the density of states close to the
Fermi surface can be neglected since the thermal energy is large
compared to the energy difference between successive Landau levels.
In any case, we note that a more detailed numerical treatment may
reveal important corrections to the estimates that follow.

Interestingly enough, there is an astrophysical environment that is
described by the above key features.  This is the interior of White
Dwarfs (WD's) which is a strongly coupled plasma of Carbon and Oxygen,
supported by a degenerate gas of electrons, similar to the iron core
of the Earth.  Hence, we adopt the formalism used for WD cooling by
axion emission in the bremsstrahlung process $e \,N(Z,A) \to e
\,N(Z,A) \,a$ \cite{Raffelt:1985nj}, in order to estimate the
geo-axion flux; $Z$ is the ionic charge and $A$ is the atomic mass.
We will ignore Primakoff \cite{Primakoff}
contributions to this flux, resulting from the
interactions of thermal photons in the plasma.  This is justified,
since the density of such photons is roughly given by $({\rm eV})^3
\sim 10^{15}$~cm$^{-3}$, which is much smaller than $n_e$ in
the core.

For a plasma with only one species of nuclei, the energy emission
rate, in axions, per unit mass is given by \cite{Raffelt:1985nj}
\beq
\eps_a = (Z^2 \alpha^2 \aae)/(A m_e^2 m_u)\,T^4 \xi(p_F),
\label{eps}
\eeq
where $m_u \simeq 1.7\times 10^{-24}$~g is the atomic mass unit and $\xi(p_F)$ is a
numerical factor which only depends on $p_F$.  Numerical calculations
relevant for WD's indicate that $\xi \simeq 1$ to a good
approximation, over a wide range of parameters in the strongly coupled
regime \cite{GR}.  We thus take $\xi \sim 1$ in our calculations.  For
geo-axion emission, we then obtain \beq \eps_a \sim 10^7 \aae \,T_3^4
\;\;{\rm erg}\, {\rm g}^{-1} {\rm s}^{-1},
\label{epsa}
\eeq
where we have set $Z= \zeff\simeq 5.4$, $A=56$, and $T_3 \equiv T/10^3$~K.  Given
a core mass density of $\rho_c \simeq 10$~g cm$^{-3}$ and $T_c \approx 5 T_3$,
we get
\beq
\Phi_a \sim 10^{37} \aae \;\;{\rm erg}\, {\rm s}^{-1},
\label{Phia}
\eeq
for the flux of geo-axions.

It is interesting to inquire how geological considerations can
constrain the estimate in Eq.~(\ref{Phia}).  As a simple criterion,
and in the spirit of analogous considerations for stellar objects, we
will demand that the rate of core-cooling $\Phi_a$ be less than that
inferred from geodynamical considerations.  This rate has been
estimated to be in the range of 100~K/Gyr$=10^{-7}$~K/yr
\cite{CB2006}.  Given that the heat capacity of the Earth's core is
estimated to be $C_\oplus \sim 10^{34}$~erg/K~\cite{CB2006}, we get
for the geological rate of core-cooling
\beq \Phi_\oplus \sim
10^{27}{\rm erg/yr}\sim 10^{19} \;\;{\rm erg}\, {\rm s}^{-1},
\label{Phie}
\eeq
in agreement with Ref.~\cite{core}.
Requiring $\Phi_a < \Phi_\oplus$ yields
\beq
\aae^\oplus \lsim 10^{-18} \quad ({\rm core}{\rm -}{\rm cooling}).
\label{aebound}
\eeq

This bound is not strong compared to those from astrophysics.  For
example, the bound from solar age is $\aae \lsim 10^{-22}$ and the one
from red giant constraints is $\aae \lsim 10^{-26}$ \cite{GR}.
However, the bound in (\ref{aebound}) is within a few orders of
magnitude of the solar. Again, we note that the bound in
(\ref{aebound}) is valid for a quasi-vacuum regime and not the extreme
stellar environments.  The closest such bounds for quasi-vacuum
environments are from $e^+ e^-$ collider experiments, and correspond
to $\aae \lsim 10^{-14}$ ($f \gsim 1000$~GeV) \cite{Amsler:2008zzb},
weaker than our geo-axion bound (\ref{aebound}) by 4 orders of
magnitude.  Next, we will examine the prospects for detecting a
geo-axion flux consistent with this bound.

The geo-axion flux $F_a$, corresponding to $\Phi_a$ in
Eq.~(\ref{Phia}), at $R_\oplus \simeq 6.4\times 10^3$~km (surface of
the Earth) is given by \beq F_a = \Phi_a/(4 \pi R_\oplus^2) \sim \aae
10^{30}\;\; {\rm eV}\,{\rm cm}^{-2} {\rm s}^{-1}.
\label{Fa}
\eeq
Assuming average axion energy
$\langle E_a \rangle \simeq 1$~eV, we get \beq \frac{d
  N_a}{d{\cal A} \,dt} \sim \aae 10^{30}\;\;{\rm cm}^{-2} {\rm s}^{-1}
\label{aflux}
\eeq for the flux of eV-axions at the surface of the
Earth\footnote{For axions, the non-radial flux also contributes, thus
  in principle Gau{\ss}' law can not be used. In this case, the
  difference compared to an exact treatment is about 5\%.}.

In principle, there are two ways to detect these axions: The obvious
choice is to exploit their coupling to electrons $\alpha_{ea}$ via the
axio-electric effect, in analogy with the photo-electric effect. The
cross section for the axio-electric effect is reduced by a factor
of~\cite{Avignone:2008uk} \beq
\frac{\alpha_{ea}}{2\alpha}\left(\frac{E_a}{m_e}\right)^2 \sim
10^{-10} \alpha_{ea} \eeq compared to the ordinary photo-electric
cross section. Using our bound derived in Eq.~(\ref{aebound}) and the
typical size of photo-electric cross sections of
$\ord{10^{-20}}\,\mathrm{cm}^2$, the resulting axio-electric cross
section is $\ord{10^{-48}}\,\mathrm{cm}^2$, which is quite small.  For
comparison typical neutrino cross sections are
$\ord{10^{-42}}\,\mathrm{cm}^2$.  Therefore, we next consider the
possibility to use magnetic axion-photon mixing, to convert axions
into photons.  We use the simple formula \cite{GR} \beq P_{a \gamma}
\simeq \left(\frac{B g_{a \gamma} L}{2}\right)^2,
\label{Pag}
\eeq
where $B$ is the strength of the transverse magnetic field along
the axion path, $g_{a \gamma}$ is the axion-photon coupling, and $L$
is the length of the magnetic region.  The above equation is valid
when the $q \ll 1/L$ with \beq q = |m_a^2 - m_\gamma^2|/(2 E_a),
\label{q}
\eeq
where $m_a$ is the axion mass and $m_\gamma$ is the photon effective
mass.

Note that for $m_a \lsim 10^{-3}$~eV, the geo-axion oscillation
length $q^{-1} \gsim 0.6$~m in vacuum with $m_\gamma = 0$.  We will
assume this mass range for the purposes of our discussion. An
obvious choice would be to consider an LHC-class magnet, such as the
one used by the CAST experiment~\cite{Andriamonje:2004hi}, with
$B=10$~T and $L=10$~m.  However, this magnet has a cross sectional
area of order 14~cm$^2$.  Given that the core of the Earth subtends
an angle of order $30^\circ$ as viewed from its surface, we see that
the CAST magnet will capture a very small portion of the relevant ``field
of view."  Thus, we have to consider other magnets of similar
strength, but larger field of view.  Fortunately, such magnets are
used in Magnetic Resonance Imaging (MRI), to carry out medical
research.  For example, the MRI machine at the University of
Illinois at Chicago has a magnetic field of 9.4~T, over length
scales of order 1~m \cite{MRI}.  Hence, we will use  $B=10$~T and
$L=1$~m, as presently accessible values, for our estimates.

The current best laboratory bound on
$g_{a\gamma}$ for nearly massless axions, derived under vacuum
conditions, was recently obtained by the GammeV
collaboration~\cite{Chou:2007zzc}:
$g_{a\gamma}<3.5\times 10^{-7}\,\mathrm{GeV}^{-1}$. With this value the
upper bound for the axion photon conversion probability is
\beq
P_{a \gamma} \simeq 3\times 10^{-12}\,.
\label{Pagnum}
\eeq
Thus, we conclude that magnetic detection is more promising than
axio-electric detection.

\begin{figure}[t!]
\includegraphics[width=\columnwidth]{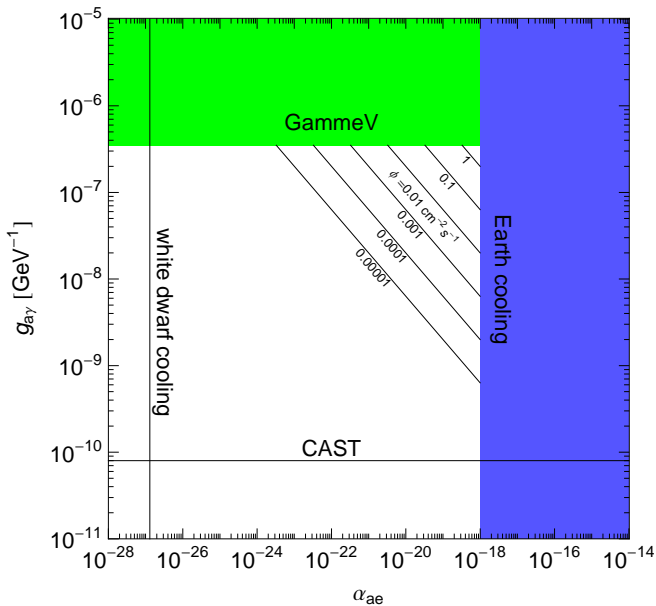}
\caption{\label{figure} The black lines show the resulting photon
  fluxes $\phi$ in units of $\mathrm{cm}^{-2}\,\mathrm{s}^{-1}$ for a
  geoscope with $L=1\,\mathrm{m}$ and $B=10\,\mathrm{T}$ as a
  function of $\alpha_{ae}$ and $g_{a\gamma}$. The gray lines show the
  current astro-physical bounds on $\alpha_{ae}$ from the cooling of
  white dwarfs~\cite{Amsler:2008zzb} and on $g_{a\gamma}$ from the
  non-observation of solar axions by CAST~\cite{Andriamonje:2004hi}.
  The colored/shaded regions indicate the parameter space excluded by
  photon regeneration~\cite{Chou:2007zzc} and the cooling of the Earth
  (this work).}
\end{figure}

Using Eq.~(\ref{aflux}), we then get
\beq \frac{d N_\gamma}{d{\cal A}
  \,dt} \sim \aae 10^{18}\;\;{\rm cm}^{-2}{\rm s}^{-1}
\label{gflux}
\eeq for the flux of converted photons in the signal.  The bound in
(\ref{aebound}) then suggests that a sensitivity to a photon flux of
order $1\;{\rm cm}^{-2}{\rm s}^{-1}$, using our reference geoscope parameters,
is required to go beyond the geodynamical constraint and
look for a signal. Modern superconducting transition edge bolometers
have demonstrated single photon counting in the near infrared with
background rates as low as $10^{-3}\,\mathrm{Hz}$~\cite{noise} and
quantum efficiencies close to unity~\cite{qe}. With such a detector
photon fluxes as small as $10^{-5}\,\mathrm{cm}^{-2}\,\mathrm{s}^{-1}$
can be detected with an integration time of $10^7\,\mathrm{s}$. Our
results on the prospects of direct search for geo-axions are
summarized in figure~\ref{figure}.

Before closing, we would like to point out a few directions for
improving our estimates.  First of all, our picture of the iron core
is quite simplified.  A more detailed treatment of electron-ion
interactions in the molten core, as well as the inner core
contribution, which was ignored here, could reveal extra
enhancements or suppressions that were left out in our analysis.
This could, in principle, require a numerical simulation of the
strongly coupled plasma (molten Fe) and the crystalline solid core.
Another issue is the possible role that the Fe $3 d$-orbital
electrons play, given that they are delocalized over a few nuclei
and may contribute to pseudo-scattering processes inside the hot Fe
medium.  Also, there could be important bound-bound and free-bound
processes that result in the emission of axions from Fe atoms at
high temperatures.  These processes have been ignored here, but
could provide contributions comparable to those we have
estimated.  In principle, more detailed geodynamical analyzes may
yield stronger bounds on non-convective energy transfer out of the
Earth's core.  This can result in tighter bounds on axion-electron
coupling $\aae$, in the regime we considered here.  Finally, our
estimate of a geoscope signal assumed an axion flux transverse to
the magnetic field. Given the angular size of the core, as viewed
through the geoscope, we expect that effective transverse field is,
on average, suppressed by roughly $\cos^2 30^\circ$, which does not
affect our conclusions, given the approximate nature of our
estimates.

In summary, we have derived estimates on possible emission of axions
from the hot core of the Earth.  Our analysis allows for possible
dependence of axion properties on non-vacuum production media, such
as astrophysical environments.  We approximated the molten core as a
strongly coupled plasma of free degenerate electrons in the
background of Fe nuclei.  We adapted the existing axion emission
estimate from a White Dwarf interior, which is a strongly coupled
plasma supported by a degenerate electron gas.  We obtained the
bound $\aae^\oplus \lsim 10^{-18}$ on the axion-electron coupling,
by considering geodynamical constraints on core-cooling rates.
Given that geo-axions would originate from a far less extreme
environment than stellar cores, our bound is nearly a vacuum bound.
Hence, our result improves existing accelerator constraints on
$\aae$, in vacuo, by 4 orders of magnitude. We also estimated the
signal strength to be expected in a dedicated search for geo-axions
using a geoscope, based on magnetic axion-photon conversion.

\acknowledgments

We would like to thank G. Khodaparast, S. King, and Y. Semertzidis
for useful discussions.  The work of H.D. is supported in part by
the United States Department of Energy under Grant Contract
DE-AC02-98CH10886.


\end{document}